\newcommand{\tb}{\ensuremath{\overline{\theta}}}

\newcommand{\tiu}{\ensuremath{J~m^{-2}~s^{-0.5}~K^{-1}}}

\newcommand{\rr}{}
\newcommand{\rrr}{}

\documentclass[authoryear,review]{elsarticle}
\journal{Icarus}

\usepackage{url}
\usepackage{natbib}
\usepackage{amsmath}
\usepackage{graphicx}
\usepackage{epsfig}
\usepackage{longtable}
\usepackage{lastpage}
\usepackage{array}

\usepackage{amssymb}

\usepackage{lineno}
\usepackage{url}

\begin{document}

\begin{frontmatter}

  \title{Determination of physical properties of the asteroid (41)
    Daphne from interferometric observations in the thermal infrared
    \tnoteref{ESOID}} 
    \tnotetext[ESOID]{Based on observations collected
    at the European Southern Observatory (ESO), Chile: ESO Program ID
    080.C-0195, VISA-Italy.}

\author[FizeauOCA]{Alexis Matter\corref{cor1}\fnref{fn2}}
\cortext[cor1]{Corresponding author.}
\fntext[fn2]{Present address: Max-Planck Institut f$\ddot{u}$r Radioastronomie, Auf
dem H$\ddot{u}$gel 69, 53121 Bonn, Germany.}
\ead{amatter@mpifr-bonn.mpg.de} 
\author[Cassiopee]{Marco Delbo}
\author[INAF]{Sebastiano Ligori}
\author[Cassiopee]{Nicolas Crouzet\fnref{fn3}}
\fntext[fn3]{Present address: Space Telescope Science Institute, 3700 San Martin Drive, Baltimore, MD 21218, USA.}
\author[Cassiopee]{Paolo Tanga}

\address[FizeauOCA]{UNS-CNRS-Observatoire de la C\^ote d'Azur,
  Laboratoire Fizeau, BP 4229 06304 Nice cedex 04, France.}
\address[Cassiopee]{UNS-CNRS-Observatoire de la C\^ote d'Azur,
  Laboratoire Cassiop\'ee, BP 4229 06304 Nice cedex 04, France.}
\address[INAF]{INAF-Osservatorio Astronomico di Torino, Strada
  Osservatorio 20, 10025 Pino Torinese, Torino, Italy.}
\end{frontmatter}

\begin{flushleft}
\vspace{2cm}
Number of pages:   \pageref{LastPage} \\
Number of tables:  3\\
Number of figures: 8\\
\end{flushleft}

\newpage
\begin{flushleft}
\begin{Large}
\vspace{1cm}
Proposed running head: Asteroid interferometry in the thermal IR.\\

\vspace{1cm}
\noindent Editorial correspondence and proofs should be directed to: \\
Alexis Matter\\
 Max-Planck Institut f$\ddot{u}$r Radioastronomie, \\
 Auf dem H$\ddot{u}$gel 69, 53121 Bonn, Germany.\\
email: amatter@mpifr-bonn.mpg.de; \\
Tel: +49 228 525 186\\
Fax: +49 228 525 229
\end{Large}
\end{flushleft}

\clearpage
\section*{Abstract}
\begin{linenumbers}
  We describe interferometric observations of the asteroid (41) Daphne
  in the thermal infrared obtained with the Mid-Infrared
  Interferometric Instrument (MIDI) and the Auxiliary Telescopes (ATs)
  of the \rr{European Southern Observatory (ESO)} Very Large Telescope
  Interferometer \rr{(VLTI)}. We derived the size and the surface
  thermal properties of (41) Daphne by means of a thermophysical model
  (TPM), which is used for the interpretation of interferometric data
  for the first time.
  From our TPM analysis, we derived a volume equivalent diameter for
  (41) Daphne of 189~km, using a non-convex 3-D shape model derived
  from optical lightcurves and adaptive optics images (B. Carry,
  private communication). On the other hand, when using the convex
  shape of \rr{Kaasalainen et al. (2002. Icarus 159, 369--395)} in our
  TPM analysis, the resulting volume equivalent diameter of (41)
  Daphne is between 194 and 209 km, depending on the surface
  roughness. 
  The shape of the asteroid is used as an \textit{a priori}
  information in our TPM analysis. No attempt is made to adjust the
  shape to the data. Only the size of the asteroid and its
  thermal parameters such as, albedo, thermal inertia and
  roughness are adjusted to the data.  We estimated our model
  systematic uncertainty to be of 4\% and of 7\% on the determination
  of the asteroid volume equivalent diameter depending on whether the
  non-convex or the convex shape is used, respectively.
%
  In terms of thermal properties, we derived a value of the surface
  thermal inertia smaller than 50~\tiu\  and preferably in the range
  between 0 and $\sim 30$~\tiu. Our TPM analysis also shows
  that Daphne has a moderate macroscopic surface
  roughness.

\vspace{3cm}
\section*{Keywords}
\noindent
  Asteroids; \sep Asteroids surfaces; \sep
  Infrared observations; \sep Data reduction techniques.


\newpage
\section{Introduction}
Information about sizes and shapes of asteroids provides essential
constraints to the history and formation processes of these bodies
\citep{2005Icar..179...63B,2009ApJ...706L.197T}. The size distribution
of the different subpopulations of asteroids and of the asteroid
dynamical families constrain the collisional evolution processes that
these bodies have experienced during their history
\citep{2005Icar..179...63B}.
\rr{Moreover, accurate determination of sizes and shapes is crucial
  to estimate volumes of asteroids, which allow one to calculate the
  bulk densities of these bodies when their masses are determined by
  some means}.
\rr{The density and internal structure are among the most important
  characteristics of asteroids, yet they are also some of the least
  constrained. When compared with the densities of meteorites - a
  partial sample of the building blocks of asteroids that survive the
  passage through the Earth's atmosphere - one can deduce the nature
  of asteroid interiors. These physical properties of asteroids
  reflect the accretional and collisional environment of the early
  solar system.}

The determination of the volumes of asteroids will be particularly
important in the next future when more asteroid masses are expected to
be accurately derived. For instance, it has been estimated that the
masses of slightly more than 100 asteroids will be determined to
better than 30\% (relative accuracy) from the gravitational
perturbations that these bodies exert on the orbits of smaller
asteroids thanks to the high accuracy astrometric measurements of the
ESA space mission Gaia \citep[launch in
2013;][]{2007A&A...472.1017M}. Yet, the volume of these bodies are not
known with accuracies small enough to allow one to calculate
meaningful densities.  \rr{The volumes of asteroids} are affected by
large \rr{errors because their true 3-dimensional shapes} are generally unknown and thus approximated by means of spheres. It can be noted, for instance that
by using a shape and a spin solution derived from lightcurves and
\rrr{mutual} occultation events for the asteroid (22) Kalliope,
\citet{2008Icar..196..578D} significantly revised its volume and thus
its density compared to previous estimates based on a sphere.
The error in the estimation of the volume can also be \rr{significant}
when large scale topographic concavities, known to be present on
asteroids \rr{\citep[see
  e.g.,][]{1999Icar..140...17T,2002Icar..155...18T}},
are approximated with flat surfaces.\\

From the size ($D$) and the \rr{absolute} magnitude of an asteroid
in the visible light ($H$), one can derive the geometric visible albedo ($p_V$)
using
the formula:
\begin{equation}
D (km) = 1329~p_V^{-1/2}~10^{-H/5}.
\label{eq.H2D}
\end{equation}
The value of the albedo is important to constrain the nature of
asteroids: it is known, for instance, that asteroids with spectra
similar to carbonaceous chondrite meteorites (types CI and CM), the
so-called C-type asteroids, have values of $p_V$ between 0.03 and 0.10
\rr{\citep{2004Icar..170..295S}}; stony (S-type) asteroids, rich in
silicates such as olivine and pyroxene have moderate values of $p_V$
(e.g., between $\sim$0.15 and $\sim$0.3)
\rr{\citep{2004Icar..170..295S}}, whereas asteroids whose reflectance
spectrum is analog to that of enstatite meteorites are known to have
in general high ($>$ 0.4) albedo values \citep[see e.g.,][and
references therein]{1989AJ.....97..580T}.  \rr{In general sizes and
  albedos of asteroids are obtained form photometric observations of
  these bodies in the thermal infrared \citep[see][for a review on the
  topic]{2002aste.conf..205H}. Models of the surface temperature
  distribution and the corresponding infrared emission are used for
  the analysis of observational data
  \citep{2002M&PS...37.1929D,2002aste.conf..205H}. In particular,
  thermophysical models (TPM) take explicitly into account the effects
  of thermal inertia, spin state, asteroid shape and surface roughness
  on the calculation of asteroids infrared emission.

  One of these parameters, the thermal inertia, a measure of the
  resistance of a material to temperature change, is particularly
  important. It is defined by $\Gamma=\sqrt{\rho \kappa c}$, where
  $\kappa$ is the thermal conductivity, $\rho$ the density and $c$ the
  specific heat. The value of the thermal inertia depends on the
  material properties \citep[see][and references therein for a table
  of thermal inertia values of some typical materials]{Mueller2007}
  and inform us about the nature of the surface regolith: a soil with
  a very low value of $\Gamma$, for instance in the range between 20
  and 50 \tiu, is covered with fine dust; an intermediate value
  (150-700 \tiu) indicates a coarser, mm- to cm-sized, regolith as
  observed on (433) Eros
  \citep{2001Natur.413..390V,2001Sci...292..484V} and (25143) Itokawa
  \citep{Yano2006}, respectively; solid rock with very little porosity
  is known to have thermal inertia values of more than 2500 \tiu\
  \citep{Jakosky1986}. The correlation between the value of $\Gamma$
  and the nature of the soil has been also demonstrated from study of
  the martian surface \citep[see e.g.,][]{2003Sci...300.2056C}.
  Moreover, because thermal inertia controls the surface temperature
  distribution of an asteroid, it affects the strength of the
  Yarkovsky effect. This is the gradual drifting of the semi major axis
  of the orbits of km-sized asteroids caused by the asymmetric (with
  respect to the direction asteroid-sun) emission of the thermal
  infrared radiation that carry momentum \citep[see][and references
  therein]{2006AREPS..34..157B}. This effect plays a role in the
  delivery of near-Earth asteroids from the main belt
  \citep{2003Icar..163..120M}, in the dispersion of asteroid families
  \citep{2004Icar..170..324N}, and it is a major source of uncertainty in the impact
  prediction estimations for potentially hazardous asteroids
  \citep{Giorgini2002,Milani2009}. Finally, accurate determination of thermal
  inertia is important in the estimation of systematic errors on
  sizes and albedos of asteroids, when these parameters are determined
  by means of simple thermal models \citep[see
  e.g.,][]{1989Icar...78..337S}.}

\smallskip As shown by \citet{2009ApJ...694.1228D}, the Very Large
Telescope Interferometer (VLTI) of the European Southern Observatory
(ESO) can be used to obtain measurements of asteroid sizes and shapes.
Generally speaking, the VLTI has the ability of measuring directly
sizes \rr{and deriving rough information about the shape of asteroids}
from measurements of the visibility (contrast) of interferometric
fringes.  The visibility is a function of the apparent angular
extension of the body along the projected interferometer
baseline. Shape features such as large concavities,
bilobed shapes and/or presence of satellites, also produce a
clear signature in the visibility. A sensitive instrument to measure
asteroid visibilities at the VLTI in the \rrr{mid-infrared} N-band (8-13
$\mu m$) is the Mid-Infrared Interferometric Instrument
\citep[MIDI;][]{2003Ap&SS.286...73L}.  The \rrr{angular} resolving power of the
interferometer depends on the length of the baseline.  VLTI baselines
vary between 16 and 130 m, with theoretical corresponding angular
resolutions between 130 and 16 mas (milliarcseconds) at 10 $\mu m$.

Interferometric observations of asteroids with other facilities, such
as the Fine Guidance Sensors (FGS) of the Hubble Space Telescope
\citep[HST,][]{2001Icar..153..451T,2002A&A...391.1123H,2003A&A...401..733T},
demonstrated the capability of the method of obtaining sizes of
asteroids and reconstructing the ellipsoids that best fit their
shape. Given the limiting magnitude of the FGS, only bright
(V$\leq$12) and large ($\sim$100 km-sized) asteroids were observed by
this program.  However, only the large VLTI baselines can overcome
these sensitivity and size limitations by extending the use of the
interferometric technique to a large number of fainter and smaller
targets. One particularly interesting feature of the MIDI instrument
is that it also measures the total (non coherent) spectral energy
distribution, $I(\lambda$), of the source in the 8-13 $\mu$m spectral
interval. This thermal infrared data can then be used as a further
constraint to derive asteroid sizes, through the application of
asteroid thermal models.  In their work, \citet{2009ApJ...694.1228D}
show the first successful interferometric observations of two
asteroids with MIDI, (234) Barbara and (951) Gaspra.

\smallskip
In this \rr{work}, we report on the continuation of the program
devoted to measurement of the physical properties of asteroids from
interferometric observations in the thermal infrared. In particular,
we obtained the first successful interferometric observations of
asteroids using the Auxiliary Telescopes (ATs) of the
VLTI. \rr{From fluxes and interferometric visibilities measured in
  the thermal infrared, we derived the size of the asteroid
  (41) Daphne, and we studied its thermal properties by means of a
  TPM. }

This \rr{work} is structured as follows: in section~\ref{s:models} we
detail the thermophysical model used for the interpretation of MIDI
data in terms of asteroid physical properties; in section~\ref{s:MIDI}
we report the observations and the data reduction process that we
adopted; \rr{in section~\ref{s:shape}, we detail the shape models that
  we used;} in section~\ref{s:results}, we give our results, followed
by a discussion in section~\ref{s:discussion}.

 \section{Modeling and analysis of MIDI observations}
\label{s:models}
MIDI is used to coherently combine the infrared light collected by two
of the four 8-m Unit Telescopes (UT) or by two of the four 1.8-m
Auxiliary Telescopes (AT) of the ESO VLTI. The two observables
measured by MIDI are the photometric flux $I(\lambda)$ and the
visibility $V(u,v)$ of the source; where $u = B_x/\lambda$ \rr{and} $v
= B_y/\lambda$ \rr{are the spatial frequencies in rad$^{-1}$ along the
  x- and y-axis, with $B_x$ and $B_y$ the components, along the two
  axis, of the interferometer's baseline, projected on the plane of
  the sky. \rrr{The x and y-axis define the coordinates on this plane.}} \rr{We recall that $V(u,v)$ is the Fourier transform of
  the brightness distribution of the source divided by
  $I(\lambda)$. The visibility can be also indicated with
  $V(\mathbf{B}/\lambda)$, where $\mathbf{B}$ is a vector of components $(B_x,B_y)$.}

\cite{2009ApJ...694.1228D} interpreted $I(\lambda)$ and $V(\mathbf{B}/\lambda)$
using simple thermal models and simple geometric models (disk of
uniform brightness and system of two disks) in order to derive the
size of (951) Gaspra and (234) Barbara. For the latter asteroid, a
disk of uniform brightness poorly reproduced the observations, whereas
a binary model provided a good fit to the data. For this reason
\cite{2009ApJ...694.1228D} speculated that Barbara has a bilobed
shape or a satellite.  When we have {\it a priori} information
about the shape and the spin state of an asteroid, TPMs can be used to
derive the size of the body and to constrain surface properties, such
as albedo and thermal inertia and
macroscopic roughness. These parameters are explicitly taken into
account in the TPM to calculate the asteroid's thermal emission, and
are adjusted until model fluxes best fit simultaneously observations
obtained at different epochs and wavelengths in the thermal
infrared. In general, these observations are measurements of the
object's disk integrated thermal infrared flux $I(\lambda)$
\rr{\citep[see e.g.,][]{2007A&A...467..737M,2009P&SS...57..259D}}.

Here, we used a TPM to calculate interferometric visibilities of
asteroids in the thermal infrared for the first time. Our procedure
consists \rrr{in generating images} of the thermal infrared emission of the
asteroid at different wavelengths as viewed by the observer and then
obtaining the model visibility and flux for each image: \rr{the model
  flux} is calculated by taking the integral of all pixels at each
wavelength, while the model visibility is calculated as the modulus of
the Fourier transform of the image, along the projected baseline
direction, divided by the flux.

The free parameters of the TPM are adjusted in order to minimize
the distance between the disk integrated flux $I'(\lambda)$ and
visibility $V'(\mathbf{B}/\lambda)$ of the model, and the corresponding
observed quantities $I(\lambda)$ and $V(\mathbf{B}/\lambda)$. As goodness of
the fit indicator, we use the reduced $\chi^2$, namely: 
  \begin{equation}  
    \overline{\chi}^2 = \frac{1}{N} 
    \left [
      \sum_{i=1}^{N_{e}}
      \sum_{j=1}^{F_i} \left
        (\frac{I_i(\lambda_j) - I'_i(\lambda_j)}{\sigma_{I_{i,j}}}
      \right )^2+ 
      \sum_{i=1}^{N_{E}}
      \sum_{j=1}^{W_i}\left(\frac{V_i(\mathbf{B}/\lambda_j) -
          V'_i(\mathbf{B}/\lambda_j)}{\sigma_{V_{i,j}}}\right)^2 
    \right ]
\label{eq:TPMchisq}
\end{equation}
where the indexes $i$ and $j$ run over the observation epochs and the
discrete samples in wavelength at which the visibility and the flux
were measured, \rr{{$N_{e}$} and {$N_{E}$}} are the
number of epochs at which flux and visibility are respectively
measured, \rr{{$F_i$} and {$W_i$}} are the number of flux and visibility samples
\rr{at the i$^{\rm th}$ epoch}, \rr{{
    $N=\sum_{i=1}^{N_{e}}N_{\lambda,I}^i+\sum_{i=1}^{N_{E}}N_{\lambda,V}^i$}}
is the total number of measurements, and $\sigma_{V_{i,j}}$ and
$\sigma_{I_{i,j}}$ are the
uncertainties on the measured visibilities and fluxes, respectively.

The physical parameters of our TPM, are:
\begin{itemize}
\item An {\it a priori} information about the shape of the body,
  described by a mesh of planar triangular facets and the spin vector
  of the asteroid. The shape and the spin vector are in general
  determined by lightcurve inversion \rr{\citep[see][for a
    review]{2002aste.conf..139K}, disk-resolved imaging \citep[from
    e.g., \textit{in-situ}, the Hubble Space Telescope, and/or
    ground-based adaptive optics observations; see
    e.g.,][]{2010Icar..205..460C}, or radar \citep[see
    e.g.,][]{2000Sci...288..836O}}.  Our implementation of the TPM
  allows non-convex shapes to be used.  \rr{Shadowing of facets to the
    observer due to the body's topography is fully taken into account.
    Mutual heating and light reflection between facets within
    topographic concavities are not modeled in this version of the
    TPM. However, their effect is of second order on the surface
    temperature determination}. The spin vector solution is given by
  the rotational phase $\phi_0$ at a reference epoch $t_0$, the
  ecliptic longitude $\lambda_0$ and latitude $\beta_0$ of the spin
  axis direction, and the rotation period $P$. Note that $t_0$ can
  be quite far in the past. As a consequence, the absolute rotational
  phase, $\phi(t)$, of the asteroid, at a more recent epoch (e.g., the
  time of VLTI observations), can be affected by a significant
  uncertainty, $\Delta_\phi$. The latter depends on the error
  $\sigma_P$ on $P$, the value of $P$, and by how far $t$ is from
  $t_0$. Since $\phi = \phi_0 + 2\pi (t-t_0) P^{-1}$, we can write
  that
  \begin{equation}
    \Delta_\phi = 2  \pi (t-t_0) P^{-2} \sigma_P, 
  \end{equation}
  neglecting the error on $\phi_0$ (which is safe - in general - to
  assume small).  When $\Delta_\phi \gtrsim 10^\circ$, then
  $\Delta_\phi$ should be treated as a free parameter of the TPM.
\item The size of the body. This is described by a factor $a$ that
  linearly scales all vertexes of the mesh. We give the size of the
  body in terms of the diameter of the sphere of equivalent volume
  $D_\vee=2 \left( \frac{3 \vee}{4 \pi} \right)^{\frac{1}{3}}$, where
  $\vee$ is the volume of the mesh.
\item The bolometric Bond's albedo $A$. This is related to $p_V$ via
  the relation: $A=p_V(0.29 + 0.684G)$, where $G$ is the slope
  parameter of the $H,G$ system of \citet{1989aste.conf..524B}.
  \rr{ Although the value reported in the MPC for
    (41) Daphne is 0.10, the use of the typical default value, namely G=0.15 \citep{1989aste.conf..524B}, does not significantly affect
    our results and we decided to keep it. }
\item The macroscopic surface roughness. This is modeled by adding
  hemispherical craters of \rrr{opening angle}, $\gamma_c$, and \rrr{surface density}, $\rho_c$. Following
  \citet{2009P&SS...57..259D}, we used here four preset combinations
  of $\gamma_c$ and $\rho_c$ spanning the range of possible values of
  surface roughness. These values of macroscopic roughness are given
  in Table \ref{tab:roughmod}, including the corresponding value of
  the mean surface slope, \tb, as defined by \citet{1984Icar...59...41H}.

\item The value of thermal inertia, which affects the temperature of
  each tile of the mesh and the temperature distribution inside
  craters.
\end{itemize}
Albedo, thermal inertia, and roughness are assumed
constant over the surface of the body.\\

{\bf [Table~\ref{tab:roughmod} here]}\\

In order to compute images of the model thermal emission, the first
step is to use the TPM to calculate the body's surface temperature
distribution, namely the temperature of each tile of the mesh and
inside craters at any given epoch $t$.
This is performed by solving the one-dimensional heat diffusion
equation for each tile of the mesh \citep[see][for more
details]{Mueller2007}. The boundary conditions of the heat diffusion
equation are given, at the surface, by the input radiative heating
from the Sun and irradiation of the heat into space, and, at depth, by
imposing a zero net heat flow towards the interior of the body.  The
position of the Sun (the heat source) with respect to each facet is
calculated by applying the inverse coordinates transformations of
\citet{2010A&A...513A..46A} in order to transform the heliocentric
position vector of the asteroid into the body's mesh reference frame:
the transformation is a function of the time $t$.  The method we adopt
to calculate the temperature distribution inside craters is given by
\citet{1998Icar..136..104E}. We do not explicitly model thermal
conduction inside craters, and here we use the approximation of
\citet{1998A&A...332.1123L}. This approximation cannot be used on the
night side. However, this is not a limitation for the present study
because our observations took place at a \rrr{moderate} solar phase
angle \rr{(see Table~\ref{tab:log1})}. Consequently, the fraction of
the night side seen by the observer was negligible.
Because of the finite thermal inertia value, the heat diffusion is not
instantaneous and the body temperature distribution depends on his
past illumination history. For this reason, the calculation of the
heat diffusion is started $\sim$ 100 rotations ($\sim$ a month) before
the observation epochs. We carefully checked that the body temperature
distribution stabilizes and \rr{is independent of the initial
  conditions}. From the knowledge of the body temperature
distribution, model fluxes are calculated assuming a wavelength
independent emissivity of 0.9, for each tile of the mesh, including
craters, in the direction towards the
observer. 
\rr{The emissivity of 0.9 assumed here is typical for the vast
  majority of silicate powders. The emissivity of these materials is
  within 10\% of 0.9 in the wavelength range between 7 and 14 $\mu m$
  \citep[see e.g.,][and references therein]{Mueller2007}.}
Then the three-dimensional mesh is projected on the plane of the sky
to create a two dimensional image of the asteroid.  \rr{In
order to have a reasonable image size, while remaining close to the
maximum resolution of the MIDI instrument}, we sampled the images
with a resolution of 4 mas/pixel (5 mas/pixels for the convex shape of
41 Daphne; see section \ref{s:results}).  Our results are robust with
respect to changes of the pixel scale. The value of each pixel is
calculated from the flux of the facet of the mesh on which the pixel
is projected to: namely it is the facet flux multiplied by the area of
the pixel and divided by the projected area of the
facet. Fig.~\ref{fig:images} shows two images of (41) Daphne obtained
along the lines described above, from the two shape models described
later.
These images are created for each wavelength.\\

{\bf [Fig. \ref{fig:images} here]}\\

\rr{Then the integrated flux and visibility of the model} are given by:
\begin{equation}
I'(\lambda_j) = a^2 \sum_{x,y} O'(x,y,\lambda_j)
\label{eq:tpmflux}
\end{equation}
\begin{align}
\nonumber V' \left( \frac{B}{ \lambda_j} \right) &=
\frac{FT(O'(\frac{x'}{a},\frac{y'}{a},\lambda_j))}{I'(\lambda_j)}=\frac{a\:\hat{O'}(\frac{a\:B}{\lambda_j})}{I'(\lambda_j)}
\\ &=\frac{a}{I'(\lambda_j)}\sum_{y'}\left( \sum_{x'}
O'(x',y',\lambda_j) \right) e^{\frac{-i2\pi a B}{N_{y'} \lambda_j}y'}
\label{eq:tpmvis}
\end{align}
where $FT$ is the Fourier Transform operator \rrr{applied to the brightness distribution of the asteroid $O'$}, $(x',y'$) is a system of
coordinates rotated by an angle {\small $PA$} with respect to the $(x,y)$ frame. The $y'$ axis thus coincides with the direction of the
baseline (see Fig.~\ref{fig:sketch}) so that we replace the baseline vector, noted $\mathbf{B}$, by its modulus, noted $B$, in the expression of the model visibility above. $N_{y'}$ is the number of pixels in the image along the $y'$ axis. In Eqs. \ref{eq:tpmflux} and \ref{eq:tpmvis}, we have
explicitly written the dependence of the flux and the visibility on the
mesh scale factor $a$. It is important to note that changing the scale
of an object by a linear factor $a$ is equivalent to multiply the
baseline length by a factor $a$, or inversely.\\

{\bf [Fig. \ref{fig:sketch} here]}\\

The free parameters of the TPM are: size (which is varied through the
scale factor $a$), thermal inertia, and macroscopic surface
roughness. The fit procedure involves calculation of $I'_i(\lambda_j)$
and $V'_i(B/\lambda_j)$ at each observing epoch and each wavelength $\lambda_j$,
for a number of discrete values of $\Gamma$ and macroscopic surface
roughness. Note that the dependence of the $\overline{\chi}^2$ on
$a$ is trivial, as it can be seen from Eqs. \ref{eq:tpmflux} and
\ref{eq:tpmvis}. Thus, the best value of $a$ for each discrete value
of $\Gamma$ and each roughness model can be found by
minimizing the Eq.~\ref{eq:TPMchisq}. Then, the location of the minimum
$\overline{\chi}^2$ as function of $\Gamma$ gives the best-fit
asteroid surface thermal inertia for each roughness model. Eventually,
the value of $a$ at $\Gamma$-minimum is used to determine the best-fit
value of $D_\vee$.

\rr{In some cases, the correction to the rotational phase
  $\Delta_\phi$ has to be treated as a free parameter of the TPM.  In
  this case, the location of the minimum $\overline{\chi}^2$, as a function of
  $\Delta_\phi$ for different roughness and thermal inertia values,
  gives the set of best-fit physical parameters for the asteroid. 
  \rrr{An optical lightcurve obtained quasi-simultaneously could help reduce $\Delta_\phi$.}}\\

\rrr{In the next section, we describe our observations of the asteroid (41)
Daphne and the related data reduction.}

\section{Observations and data reduction}
\label{s:MIDI}
The observations of (41) Daphne were carried out in service mode, on 2008
March 12 and 14 . We adopted the typical observing sequence with
MIDI, which is extensively described by \citet{2003Ap&SS.286...85P}.
Using two ATs in the E0-G0 baseline configuration ($B=16$~m), we
acquired four visibility observations for (41) Daphne, two on March
12 and two on March 14. Table~\ref{tab:log1} reports the observational
circumstances. The telescopes and the delay lines of the
interferometer were tracked at the rates predicted from the
ephemerides of the target.\\

{\bf [Table.~\ref{tab:log1} here]}\\

Our observations included a mid-infrared photometric and
interferometric calibrator chosen from the ESO database, namely
HD123139. The absolutely calibrated infrared spectrum of the calibration
star was taken from \rr{\citet{1999AJ....117.2434C}}. \rr{Extraction
  and calibration of the visibility measurements of (41) Daphne were
  performed using the same method as \citet{2009ApJ...694.1228D}.}
The flux and visibility measurements of (41) Daphne are shown in
Fig~\ref{fig:flux_vis_bestfit_convex}~(or identically in
Fig~\ref{fig:flux_vis_bestfit_concave}).  \rr{The estimation of error
  bars constitutes a difficult issue when reducing MIDI data.  In the
  most common case, when the `high-sensitivity' mode is used, the
  photometry is acquired about 3 to 6 minutes after the fringes are
  recorded.  Therefore, the measured value of $I(\lambda)$ does not
  correspond to the flux of the source at the time of the fringes
  recording. This leads to typical uncertainties of about 10 to
  15\% \citep[see][for more details]{2007NewAR..51..666C}. This error
  strongly depends on the atmospheric conditions during the night. An
  estimation of the error bars can thus be obtained by computing the
  RMS of the visibilities values of several calibrators observed
  closely to the source (or identically several observations of the
  same calibrator). We used this procedure for the two first
  visibilities, and considered a typical relative uncertainty of 15\%
  for the two others since only one calibrator was observed during the
  second night.}
 
  {\bf [Fig.~\ref{fig:flux_vis_bestfit_convex} and 
    Fig.~\ref{fig:flux_vis_bestfit_concave} here]} \\

\rr{In addition to the interferometric observations, we obtained an
  optical lightcurve of (41) Daphne, in order to better constrain the
  rotational phase at the epoch of the VLTI observations. In October
  2009, during the tests of the 0.4 m telescope for the Antarctic
  Search for Transiting ExoPlanets (ASTEP) project
  \citep{astep} at the Observatoire de la C\^ote d'Azur
  in Nice, France, we performed time-resolved CCD photometric
  observations of this asteroid in the visible. Our observations
  spanned a period of about 3 h.  We used the technique of
  differential photometry, which makes use of stars present in the
  same CCD frame as the target, to allow for accurate removal of
  systematic effects such as atmospheric extinction variability. The
  telescope was tracked at the sidereal rate. The reduction of the CCD
  frames consisted in the conventional \rrr{bias} removal and flat-fielding,
  which was performed by using \rrr{bias} and flat calibration frames.  The
  corresponding lightcurve is shown in Fig.~\ref{fig:djamel}, where
  the times of the
  observations were light-travel subtracted.}\\
 
{\bf [Fig.~\ref{fig:djamel} here]} \\

\rr{In the next section we describe the different shape models used
  for (41) Daphne, and how the TPM analysis was performed using this
  information).}

\section{Shape models and TPM analysis}
\label{s:shape}
For sake of comparison and evaluation of the results variability, two
shape models were used in this study:
\begin{enumerate}
\item a convex mesh was downloaded (March 2010) from the Database of
  Asteroid Models from Inversion Techniques or DAMIT \citep[see][and
  \url{http://astro.troja.mff.cuni.cz/projects/asteroids3D/web.php?
    page=project_main_page}]{2010A&A...513A..46A}. This is a database
  of three-dimensional models of asteroids derived by solving the
  inverse problem of determining the object's shape and its rotational
  state from optical lightcurves. \rrr{It also lists some models
    derived from a combination of adaptive optics images and
    lightcurves}.  The convex shape of (41) Daphne contains 1022
  vertexes and 2040 triangular planar facets. The pole solution is
  also given: $\lambda_0$=198$^o$ $\beta_0$=-32$^o$, P=5.98798 h,
  $\phi_0$=0$^o$ at the reference epoch $t_0$=2444771.79382 (JD);

\item the second shape model of (41) Daphne, \rrr{provided by B. Carry (private communication)\footnote{\label{benoit}See also his PHD thesis manuscript available at \url{http://benoit.carry.free.fr/} for further information.}}, is non-convex and was derived from the
  KOALA method
  \rr{\citep{2010Icar..205..460C,2010arXiv1001.2438K}}. We recall here that this method allows one to derive the size and the shape of an asteroid from the inversion of data obtained by different observational techniques: namely, photometric lightcurves and adaptive optics images in the case of (41) Daphne.  The
  corresponding pole solution is: $\lambda_0$=198$^o$
  $\beta_0$=-31$^o$, P=5.987980 h, $\phi_0$=0$^o$ at the reference
  epoch $t_0$=~2444771.79382 (JD).
\end{enumerate}
\rr{Note that, while the size of the first shape model is arbitrary,
  the non-convex shape, \rrr{derived from the KOALA method, has a volume equivalent diameter of $D_\vee=185\pm5$~
  km. We call it `the nominal size' in the following}.}
The pole solution of the convex model of (41) Daphne was
  obtained from inversion of a set of 23 optical lightcurves obtained
  in the period 1976 - 1988. Assuming $t_0$ =
  December 21, 1988 and $ \sigma_P$=0.00001 hours, we thus find that
  $\Delta_\phi \sim 17^{\circ}$ at the epoch of our VLTI observations,
  given the rotation period of 5.98798 hours for this object. As a
  consequence, we attempted to better constrain the rotational phase
  of (41) Daphne using our recently obtained lightcurve shown in
  Fig.~\ref{fig:djamel}.
  To do so, we calculated synthetic visible lightcurves using the
  convex shape of (41) Daphne and attempted to fit the model to the
  observed lightcurve using the value of $\Delta_\phi$ as a free
  parameter. A geometric scattering model was adopted: namely the
  visible reflected flux from each facet is proportional to the
  projected area of the facet, and it is required that the facet is
  also illuminated by the sun. Fig.~\ref{fig:djamel} also shows the
  best fit model lightcurve that we obtained using the convex model
  for a correction value of $\Delta_\phi=(-0.2\pm0.3)^\circ$.  \rr{The
    correction to the nominal rotational phase is thus negligible.\\}
  \rr{For the non-convex model, given that the KOALA model uses recent
    observations from 2008 March 28, and that the period is more
    accurate (by an order of magnitude), the error on the rotational
    phase given by the pole solution is negligible as well. As a
    consequence, for the two shape models, the rotational phase was
    not treated as a free parameter.}
  The TPM was run for each roughness model (see
  Table~\ref{tab:roughmod}), and each value of thermal inertia, namely 0, 5, 10,
  25, 50, 75, 100, 150, 200, 300, 400 and 500 \tiu.  Then the fit
  procedure described in section~\ref{s:models} was applied to the
  measured fluxes and visibilities.

  In the next section we describe and discuss the results obtained
  from the application of the TPM to the observed visibilities and
  fluxes of (41) Daphne.

\section{Results}
\label{s:results}
\subsection{Convex shape model}
\rr{Fig.~\ref{fig:chisq_convex} shows our best-fit estimator
  $\overline{\chi}^2$ as a function
  of $\Gamma$ for the four different roughness models, 
  in the case of the convex shape.}\\

{\bf [Fig. \ref{fig:chisq_convex} here]}\\

\noindent \rr{We note that a surface with a low or no
  macroscopic roughness and a value of thermal inertia $<100$~\tiu\
  gives the best fit to the observations. In particular the minima of
  the `no roughness' and `low roughness' models are at $\Gamma$=48 and
  8~\tiu, respectively. The corresponding values of $D_\vee$ are 209
  and 194 km, with associated $p_V$ values of 0.057 and 0.067,
  respectively.}
\rr{In Fig.~\ref{fig:flux_vis_bestfit_convex} we plot the visibility
  and flux of the best-fit model (`no roughness', $\Gamma=48$~\tiu,
  $D_\vee=209$~km, $p_V=0.057$), in addition to the measured fluxes
  and visibilities of (41) Daphne. We note that our model represents
  well the observed flux except for the third observing epoch, where
  the flux of the TPM is greater than the measured one (by roughly
  20\% at 13~$\mu$m).  This `offset-like' mismatch may likely come
  from an underestimation of the source flux by MIDI, especially
  around 13~$\mu$m where the signal to noise ratio is generally quite
  low \citep{2007NewAR..51..666C}. The photometry measurement of the
  source may thus be degraded by a bad estimation and suppression of
  the thermal background (and its fluctuations), which is dominant in
  the mid-infrared \citep[see e.g.,][]{2003EAS.....6..127P}.  Moreover,
  the underestimation of the flux at the third epoch implies a similar
  offset-like mismatch between the corresponding measured and
  calculated visibilities at the same epoch.}

\rr{In order to estimate the statistical uncertainty affecting the fit
  parameters $\Gamma$ and $D_\vee$, a Monte-Carlo analysis was
  performed.
  To this end, 200 normally distributed flux and visibility values per
  observation were generated at each wavelength, with average and
  standard deviation matching the measured fluxes and visibilities
  within their respective $1-\sigma$ uncertainty. For each set of fluxes and
  visibilities, a fit of the model, as described in section
  \ref{s:models}, was performed. Then, we took the standard deviation
  of the $\Gamma$ and $D_\vee$ values at the minimum
  $\overline{\chi}^2$ as the $1-\sigma$ uncertainty on our best fit values of
  thermal inertia and volume equivalent diameter.  As a result we find
  $\Gamma$ = 48$\pm$5~\tiu~and $D_\vee$ = 209$\pm$1 km as the best fit
  solution for a model using the convex shape without roughness; and
  $\Gamma$ = 8$\pm$8~\tiu~and $D_\vee$ = 194$\pm$2 km as the best fit
  solution for a model using the convex shape with a low roughness.}

\subsection{Non-convex shape model} 
\rr{Figure~\ref{fig:chisq_concave} shows our 
best-fit estimator $\overline{\chi}^2$ 
of the TPM, calculated in the case of the non-convex shape.} \\


{\bf [Fig. \ref{fig:chisq_concave} here]}\\

\noindent \rr{In this case, a model with a low or medium roughness and a
  thermal inertia $<100$~\tiu~gives the best fit to the
  observations. In particular, the minima of the `low roughness' and
  `medium roughness' models are at $\Gamma$=9 and 0 ~\tiu,
  respectively. The corresponding value of $D_\vee$ are 189 km and
  182 km, 
  respectively, with $p_V$= 0.07.  The solution with $\Gamma$=0~\tiu~
  is non physical and thus we reject it.
  Although 100 km-sized Main Belt asteroids are
  known to have low thermal inertia, their values are in general
  larger than 10~\tiu~\citep[see e.g.,][and references
  therein]{1998A&A...338..340M}. We also used the Monte-Carlo
  procedure described above to estimate the uncertainties on the fit
  parameters. These uncertainties result to be of 1~\tiu~for the
  thermal inertia and 1 km for the volume equivalent diameter.  We
  show in Fig.~\ref{fig:flux_vis_bestfit_concave} the measured fluxes
  and visibilities of (41) Daphne along with the best fit of the
  non-convex case (`low roughness', $\Gamma=9$~\tiu, $D_\vee=189$~km,
  $p_V=0.070$).  Our model, when used with the non-convex shape, also
  provides a good match to the observed flux. We note that the flux
  prediction at the third epoch is as well higher than the measured
  one. This effect may also be due to an underestimation of the source
  flux by MIDI, as described above. Moreover, the non-convex model seems
  to better fit the measured visibilities than the convex
  one. 
  This result is consistent with the fact that the non-convex shape
  model corresponds to a more realistic and detailed representation of
  (41) Daphne, as previously indicated by B. Carry (private
  communication)$^{\rm \ref{benoit}}$ who better reproduced an
  occultation profile of (41) Daphne with this shape model. We can
  note however a slight mismatch between the model predictions and the
  measurements at the fourth epoch. The mismatch is located around
  $\frac{B}{\lambda}\approx 6$~as$^{-1}$ (or $\lambda\approx 9.6
  \mu$m) i.e. close to the atmospheric ozone absorption feature. This
  time-dependent absorption feature can cause fluctuations and lower
  the signal to noise ratio on both the coherent flux and the
  photometry. Due to its relative variability, this feature is
  frequently imperfectly removed during the reduction process of
  interferometric observations \citep[see
  e.g.,][]{2004SPIE.5491..588T}.}
All the results are summarized in Table~\ref{tab:TPM}.\\

{\bf [Table~\ref{tab:TPM} here]}\\

\section{Discussion}
\label{s:discussion}
\subsection{Size and geometric visible albedo} 
The best-fit value of $D_\vee$ obtained from our TPM analysis of MIDI
data using the non-convex shape model, namely $189\pm1$~km, presents a
discrepancy with the nominal value ($185\pm5$~km) of about 4~km. This
is about 4 times greater than the statistical uncertainties implied by
the photometric and visibilities measurements and estimated via our
Monte Carlo method.  However, this discrepancy represents a relative
accuracy of 2\% on the diameter that appears remarkably good for a
typical size determination of asteroids.
If the two equally plausible solutions in terms of our best-fit
indicator, i.e. low roughness - $\Gamma$=9~\tiu\ and medium roughness -
$\Gamma$=0~\tiu, are considered in order to estimate the model error
on the value of $D_\vee$, we found that the latter is 7 km i.e. 4\% in
relative accuracy.

When the convex model is used as an \textit{a priori} shape for the
TPM, the value of $D_\vee$ resulting from the analysis of MIDI
data is between 194 and 209~km (with a mean value of about 200~km),
depending on whether a low-roughness or a no-roughness model,
respectively, is assumed for the surface of Daphne. Because both
solutions are equally good in terms of our best-fit estimator, it is
clear that the systematic model uncertainty dominates the final size
determination. This systematic relative uncertainty is of the order of
7\% on the diameter.

It is also evident that the best fit values of the equivalent volume
diameter obtained when using the convex shape are larger than the
nominal value of $D_\vee$ of the non convex shape. This is because the
latter has a smaller volume (due to the presence of
topographic concavities) for a similar projected size on the plane of
the sky. This volume overestimation when the convex shape is used,
clearly indicates that the imposed
condition of convexity, imposed by the lightcurve inversion technique
\citep[see][]{2001Icar..153...24K}, introduces a significant systematic
bias on the size when large concavities are present on the asteroid surface.

The geometric visible albedo values derived from the TPM with the
convex and non-convex shapes are in the range between 0.05 and 0.07.
This is in agreement with the C-type taxonomic classification of
(41) Daphne.

\subsection{Thermal properties} 
\rr{We obtained for the first time good constraints on the
  determination of macroscopic roughness for both shape models,
  although the results slightly differ. While a very low roughness is
  preferred in the convex shape case, a low or moderate roughness
  gives the best-fit in the case of the non-convex one.}
\rr{We estimate that the corresponding mean surface slope values should lie between 10 and 29$^\circ$ for (41) Daphne. A high macroscopic
  roughness is discarded, which is quite surprising given the
  important surface roughness expected for large asteroids.
  Large main-belt asteroids such as (1) Ceres and
  (2) Pallas are expected to have rough surfaces as found by
  \citet{1989Icar...78..337S} from the analysis of the thermal
  infrared emission of these bodies.  It is also known that the
  observed zero-phase thermal emission of the Moon is well reproduced
  by a rough-surface model, i.e.  $\rho_c=0.64$ and
  $\gamma_c=90^\circ$ \citep{1990Icar...83...27S}}. The possibility to
constrain the macroscopic roughness in the thermal infrared is very
interesting because it allows, in principle, a more accurate
determination of the asteroids thermal inertia from TPM modeling.

\rr{Our TPM analysis indicates that (41) Daphne has a thermal inertia
  value certainly smaller than 50~\tiu, and likely in the range
  between 10 and $\sim 30$~\tiu, indicating that a layer of very fine regolith covers its surface. This is in agreement with the thermal inertia values already measured on main-belt asteroids larger than 100~km in diameter. Indeed, from mid- and far-infrared observations, \citet{1998A&A...338..340M} derived very low thermal inertia values, between 5 and 25~\tiu, for some large Main Belt asteroids including (1) Ceres, (2) Pallas, (3) Juno, (4) Vesta, and (532) Herculina. Moreover, \citet{2004A&A...418..347M} derived a thermal inertia of about 15~\tiu for (65) Cybele. More recently, from the direct measurement of the cooling during shadowing events in a binary system, \citet{2010Icar..205..505M} measured a thermal inertia of $20\pm15$~\tiu for the large binary Trojan ($D\approx100$~km), (617) Patroclus.}\\

Several works were recently devoted to the determination of the
thermal inertia value of the asteroids surface from TPM analysis of
disk integrated thermal infrared data \rr{\citep[see e.g.,][and
  references
  therein]{2007Icar..190..236D,2009P&SS...57..259D,Mueller2007}}. In
general, these works clearly showed that a good fit of the model to
the observations can be obtained by using any roughness model. Because
the determination of the value of the thermal inertia is a function of
the roughness model adopted, any constraint on the latter parameter
will allow improving the determination of thermal inertia.
\rr{In order to show the effect of the visibility on the determination of 
a simultaneous solution for both thermal inertia and surface 
  roughness, we also performed a TPM analysis of the disk integrated 
  thermal infrared data of (41) Daphne, using the non-convex shape 
   and neglecting the visibility measurements.}\\

{\bf [Fig.~\ref{fig:chisq_concave_flux} here]}\\
 
\rr{As it can be seen from Fig.~\ref{fig:chisq_concave_flux}, any
  roughness model and any value of the thermal inertia give a value of 
  $\overline\chi^2$ smaller than 1, which implies that any value of these
  parameters allows the TPM to fit equally well the infrared flux. In
  order to reduce this degeneracy, it is known that thermal infrared
  photometric observations are needed at different illuminations and
  viewing geometries in order to constrain the asteroids thermal
  properties from a TPM modeling of photometric observations only.}

\section{Conclusion}
We have obtained the first successful interferometric observations of
asteroids in the thermal infrared using the ATs of the ESO VLTI. 
We observed the asteroid (41) Daphne using the MIDI instrument and the 16m-long baseline E0-G0.

We developed a thermophysical model (TPM) for the analysis of
interferometric observations of asteroids in the thermal
infrared, with the aim of deriving information about size and thermal 
properties.

We applied our TPM to the MIDI observations of (41) Daphne: our
results indicate that Daphne has a volume equivalent diameter between
194 and 209 km, depending on the surface roughness, if a convex shape model derived from lightcurve inversion
is used as an {\it a priori} constraint on the shape of the asteroid.
\rr{\rrr{Since the nominal size, attached to the non-convex KOALA model, is
  $185\pm5$~km (B. Carry, private communication)$^{\rm \ref{benoit}}$}, our results confirm that the assumption of convexity
  introduces a systematic bias on the size determination when
  important concavities are present on the asteroid's surface.  In
  contrast, if the non-convex shape model is used, the volume
  equivalent diameter \rrr{obtained from the TPM} is 189$\pm$1 km, i.e. very close to the 
  nominal value of $185\pm5$~km. We estimated our model
  systematic uncertainty to be of 4\% and of 7\% on the determination
  of the asteroid volume equivalent diameter depending on whether the
  non-convex or the convex shape is used, respectively.} 
   
Our TPM analysis also showed that the macroscopic surface roughness
can be constrained by interferometry, thanks to the angular resolving power offered by the VLTI and which allows to resolve the temperature distribution on the asteroid surface.
In particular, using both shape models of (41) Daphne, we found 
a moderate to low roughness (see Table \ref{tab:roughmod});
`high roughness' models are discarded by our analysis.  With such a
constraint on the macroscopic roughness, the TPM results indicate a
very low thermal inertia for (41) Daphne, certainly smaller than
50~\tiu. This confirmed previous results indicating that the surface
of asteroids with sizes larger than 100 km has a low thermal inertia.
As shown by this work, the possibility of constraining the macroscopic
roughness is important in the prospect of an accurate modelling of the
thermal infrared emission of asteroids, and especially thermal inertia
determination \citep[see also][]{Mueller2007,2009P&SS...57..259D}.

\end{linenumbers}
\subsection*{Acknowledgments}
We would like to thank the staff and the Science Archive Operation of
the European Southern Observatory (ESO) for their support in the
acquisition of the data. The comments from the referee Benoit Carry
and an anonymous referee were extremely helpful in revising this work
and allowed us to improve significantly the manuscript.

\newpage
\bibliographystyle{model2-names}
\bibliography{astero}

\newpage
\section*{Tables and Table Captions}

{

\begin{table}[h]
\centering
\begin{tabular}{lccc}
\hline
Roughness model      & $\gamma_c$         & $\rho_c$ & \tb \\
\hline
  {no roughness}     & {$0^\circ$}  & {0.0} & {0$^\circ$} \\
  {low roughness}    & {$45^\circ$} & {0.5} & {10$^\circ$} \\
  {medium roughness} & {$68^\circ$} & {0.8} & {29$^\circ$} \\
  {high roughness}   & {$90^\circ$} & {1.0} & {58$^\circ$}\\
\hline
\end{tabular}
\caption{The four roughness models used in the application of the 
  TPM to the MIDI data; \rr{$\gamma_c$ and $\rho_c$  
    respectively correspond to the crater opening angle and the crater density, 
    while}  \tb~is the corresponding mean surface slope according 
  to the parameterization introduced by \citet{1984Icar...59...41H}
  \citep[see text and also][for further details]{2007Icar..190..236D}.}
\label{tab:roughmod}
\end{table}

 
 \begin{table}[h]
 \centering
 \begin{tabular}{cccccccc}
 \hline
{Date}&{UT}&{$r$ (AU)}&{$\Delta$ (AU)}&{$\alpha$ (deg)}&{$B$ (m)}&{$PA$ (deg)}&{Tag}\\
 \hline
        {2008-03-12}& {05:19:14}& {2.0866}& {1.1920}& {15.8}& {13.6}&{66.6}&{D$_1$}\\
        {2008-03-12}& {06:24:35}& {2.0864}& {1.1917}& {15.8}& {15.4}&{70.6}&{D$_2$}\\
        {2008-03-14}& {04:19:27}& {2.0834}& {1.1769}& {15.0}& {11.6}&{61.6}&{D$_3$}\\
        {2008-03-14}& {04:32:48}& {2.0834}& {1.1769}& {15.0}& {12.2}&{63.4}&{D$_4$}\\
 \hline
 \end{tabular}
 \caption{{\rr{Observational circumstances and interferometric parameters of the 
(41) Daphne observations. $r$ and $\Delta$ are the heliocentric and geocentric 
distances, respectively, while $\alpha$ is the solar phase angle. $B$ and $PA$ 
are respectively the length and the position angle of the baseline projected 
on sky. The last column contains a tag associated with each observation.}}}
 \label{tab:log1}
 \end{table}

\begin{table}
\centering 
\begin{tabular}{cccccc}
\hline
{Shape}& {Roughness model} & {$\overline{\chi}^2$}&{$\Gamma$}&{D$_\vee$}& {$p_V$} \\ 
       &                   &                      &(\tiu)& (km) &\\
\hline
Convex &No roughness & $3.5\pm0.3$& $48\pm6$     & {$209\pm1$}&  {$0.057\pm0.009$}\\
       &Low roughness& $3.7\pm0.3$& $8^{+10}_{-8}$ & {$194\pm2$}&  {$0.067\pm0.011$}        \\
Non-convex&Low roughness&$3.4\pm0.3$ & $9\pm1$ & $189\pm1$  & $0.070\pm0.011$ \\
\hline    
\end{tabular} 
\caption{Results of the determination of physical properties of the 
  asteroid (41) Daphne, using the TPM. The $\overline{\chi}^2$ is 
  our best-fit estimator as described by Eq~\ref{eq:TPMchisq}; 
  $\Gamma$ is the thermal inertia, $D_\vee$ is the spherical volume 
  equivalent diameter, and $p_V$ is the geometric visible albedo. The errors are within 1-$\sigma$. 
}
\label{tab:TPM}
\end{table}

\clearpage

\section*{Figure captions and figures}

{\bf Fig~\ref{fig:images} caption :}\\
Image of the asteroid (41) Daphne created from the TPM, using the convex shape model (left image) and the non-convex one (right image). The gray level is proportional to the emitted thermal infrared flux.

{\bf Fig~\ref{fig:sketch} caption :}\\
Illustration of the geometric parameters involved in the calculation of the synthetic visibility and flux from a TPM image.

{\bf Fig~\ref{fig:flux_vis_bestfit_convex} caption :}\\
Left panels: measured thermal infrared fluxes (with error bars) between 8 and 13 $\mu$m of (41) Daphne, and the corresponding  best-fit synthetic infrared fluxes (solid lines) derived from the TPM in the case of the convex shape; right panels: measured interferometric visibilities plotted, in N band, as a function of angular frequency, and the corresponding synthetic visibilities of the TPM (solid lines). The best-fit model represented here is : `no roughness', $\Gamma=48$~\tiu. The tags $D_1$, $D_2$, $D_3$ and $D_4$ indicate the observing epoch in the chronological order (see Table~\ref{tab:log1}).

{\bf Fig~\ref{fig:flux_vis_bestfit_concave} caption :}\\
As for Fig. \ref{fig:flux_vis_bestfit_convex}, but the best-fit model represented here, in the case of the non-convex shape, is : `low roughness', $\Gamma=9$~\tiu. The tags $D_1$, $D_2$, $D_3$ and $D_4$ indicate the observing epoch in the chronological order (see Table~\ref{tab:log1}).

{\bf Fig~\ref{fig:djamel} caption :}\\
Diamonds: lightcurve obtained from CCD photometric observations in the visible of (41) Daphne in October 2009 during the tests of the ASTEP 0.4 m telescope at the Observatoire de la C\^ote d'Azur, Nice, France. These observations spanned a period of about 3
hours. Solid curve: corresponding best-fit lightcurve from the \rrr{convex} model.

{\bf Fig~\ref{fig:chisq_convex} caption :}\\
Plot of $\overline{\chi}^2$ (see Eq.~\ref{eq:TPMchisq}), calculated from the TPM in the case of the convex shape, as a function of thermal inertia $\Gamma$, for the four roughness models (see Table~\ref{tab:roughmod}).

{\bf Fig~\ref{fig:chisq_concave} caption :}\\
Plot of $\overline{\chi}^2$ (see Eq.~\ref{eq:TPMchisq}), calculated from the TPM in the case of the non-convex shape, as a function of thermal inertia $\Gamma$, for the four roughness models (see Table~\ref{tab:roughmod}).

{\bf Fig~\ref{fig:chisq_concave_flux} caption :}\\
Plot of $\overline{\chi}^2$ (see Eq.~\ref{eq:TPMchisq}) calculated from the TPM, using only the flux measurements of (41) Daphne. This is represented in the case of the non-convex shape, as a function of thermal inertia $\Gamma$, for the four roughness models (see Table~\ref{tab:roughmod}).

\newpage
\begin{figure}
 \centering
 \includegraphics[scale=0.45]{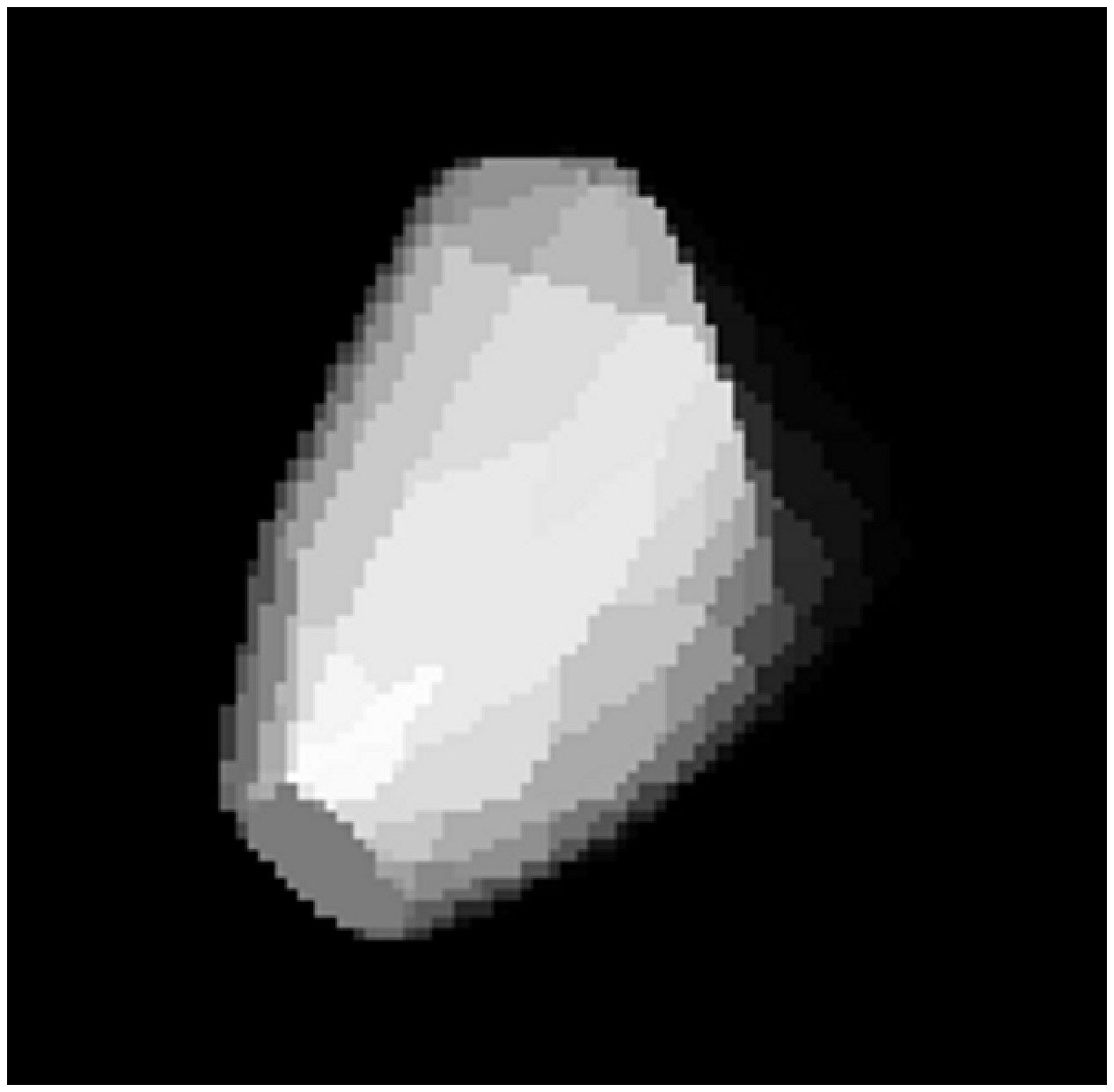} \hfill
 \includegraphics[scale=0.45]{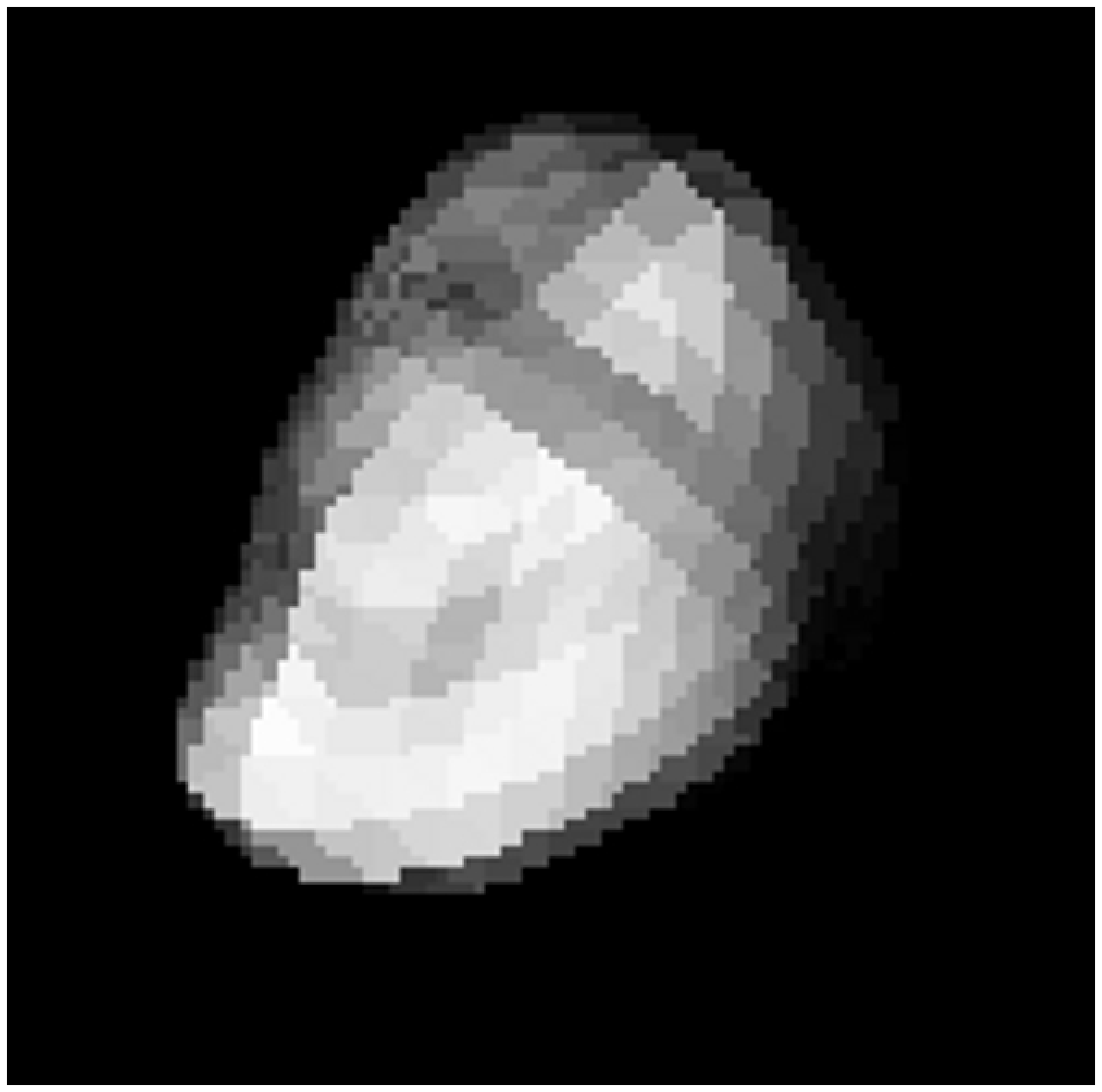}
 \caption{}
 \label{fig:images}
\end{figure}

\newpage
\begin{figure}
 \centering
 \includegraphics[width=130mm]{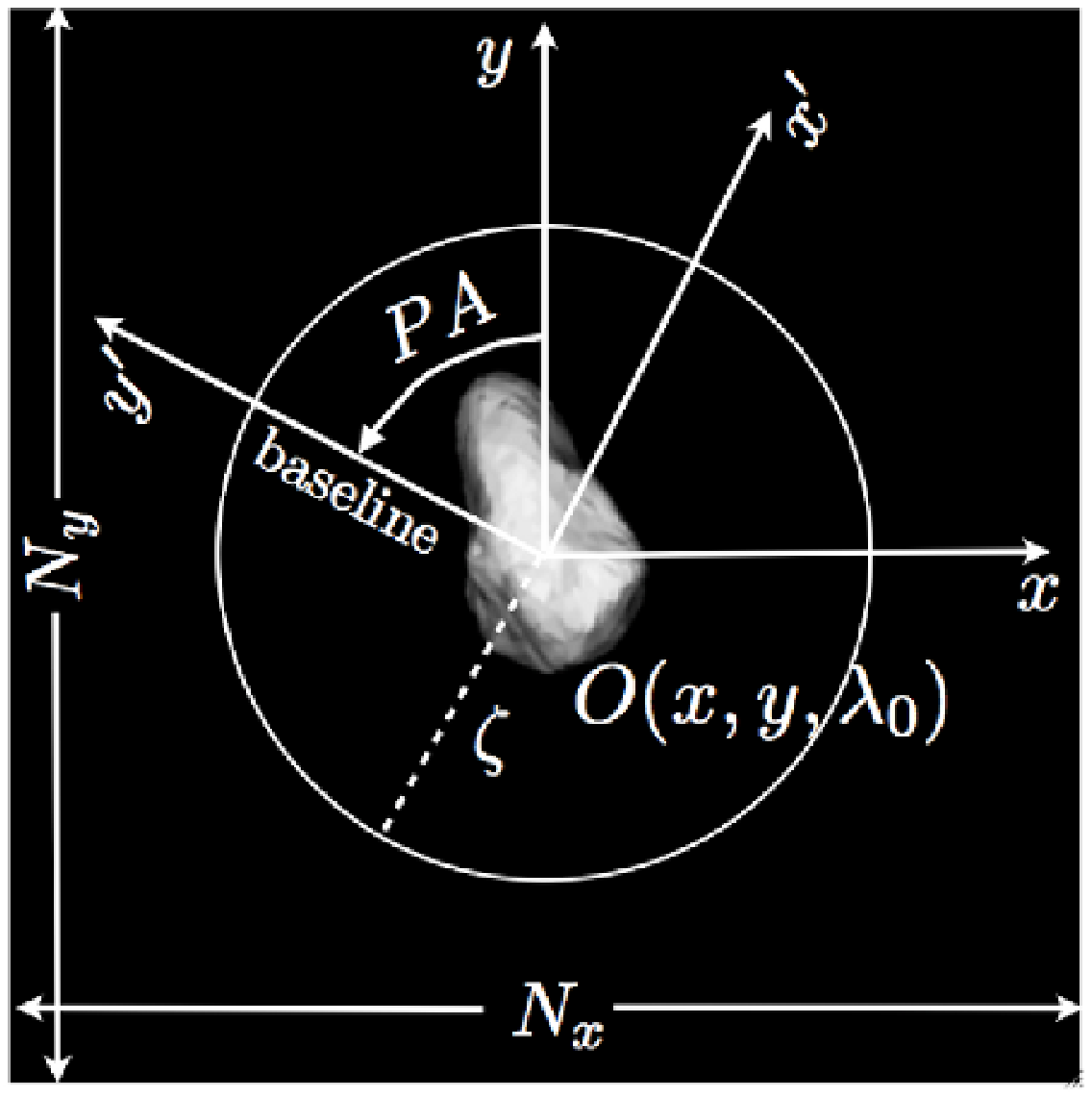}
 \caption{}
 \label{fig:sketch}
\end{figure} 

\newpage
\begin{figure}
\centering
 \includegraphics[width=150mm,height=150mm]{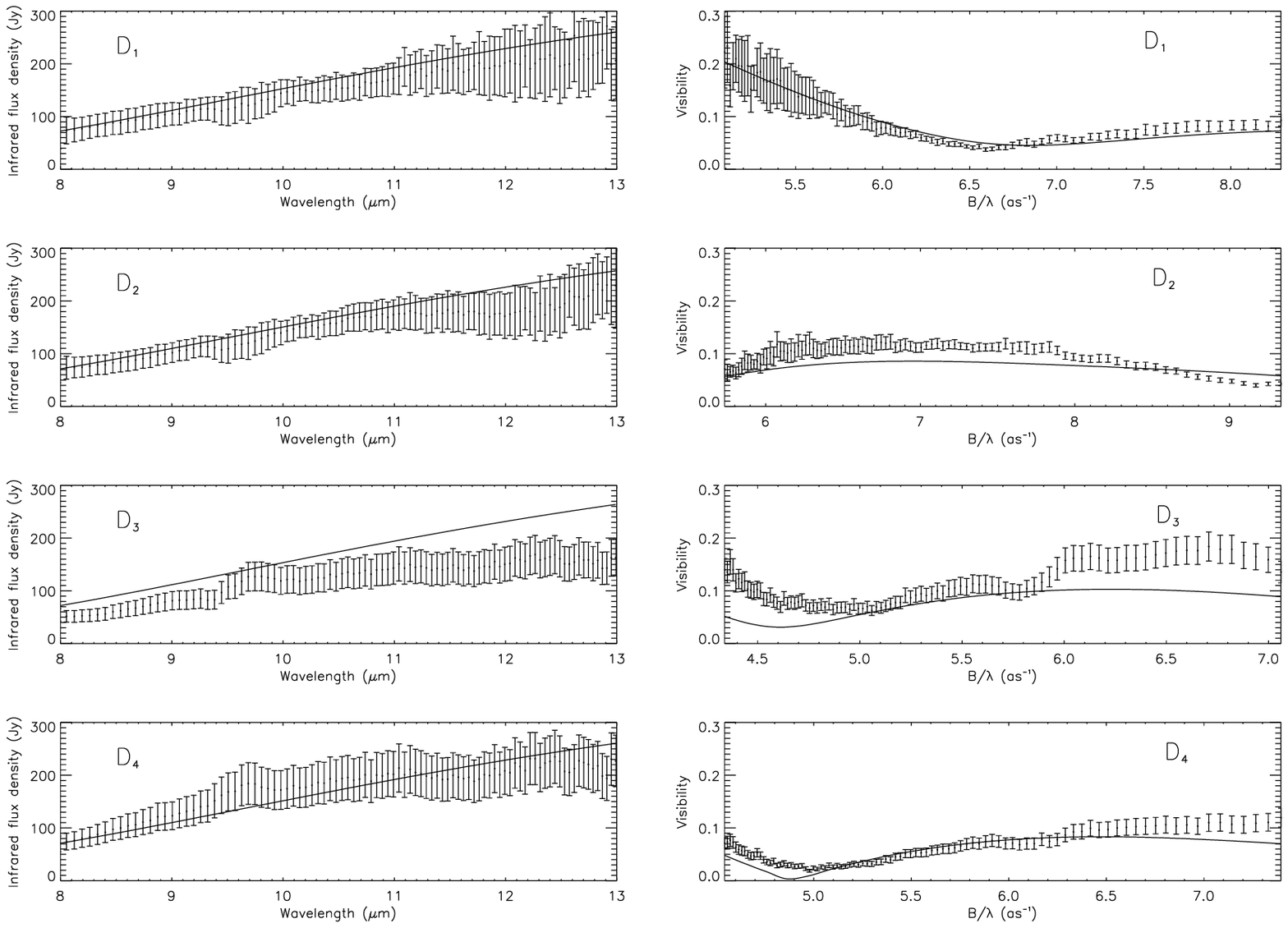}
 \caption{}
 \label{fig:flux_vis_bestfit_convex}
\end{figure} 

\newpage
\begin{figure}
 \begin{center}
 \includegraphics[width=150mm,height=150mm]{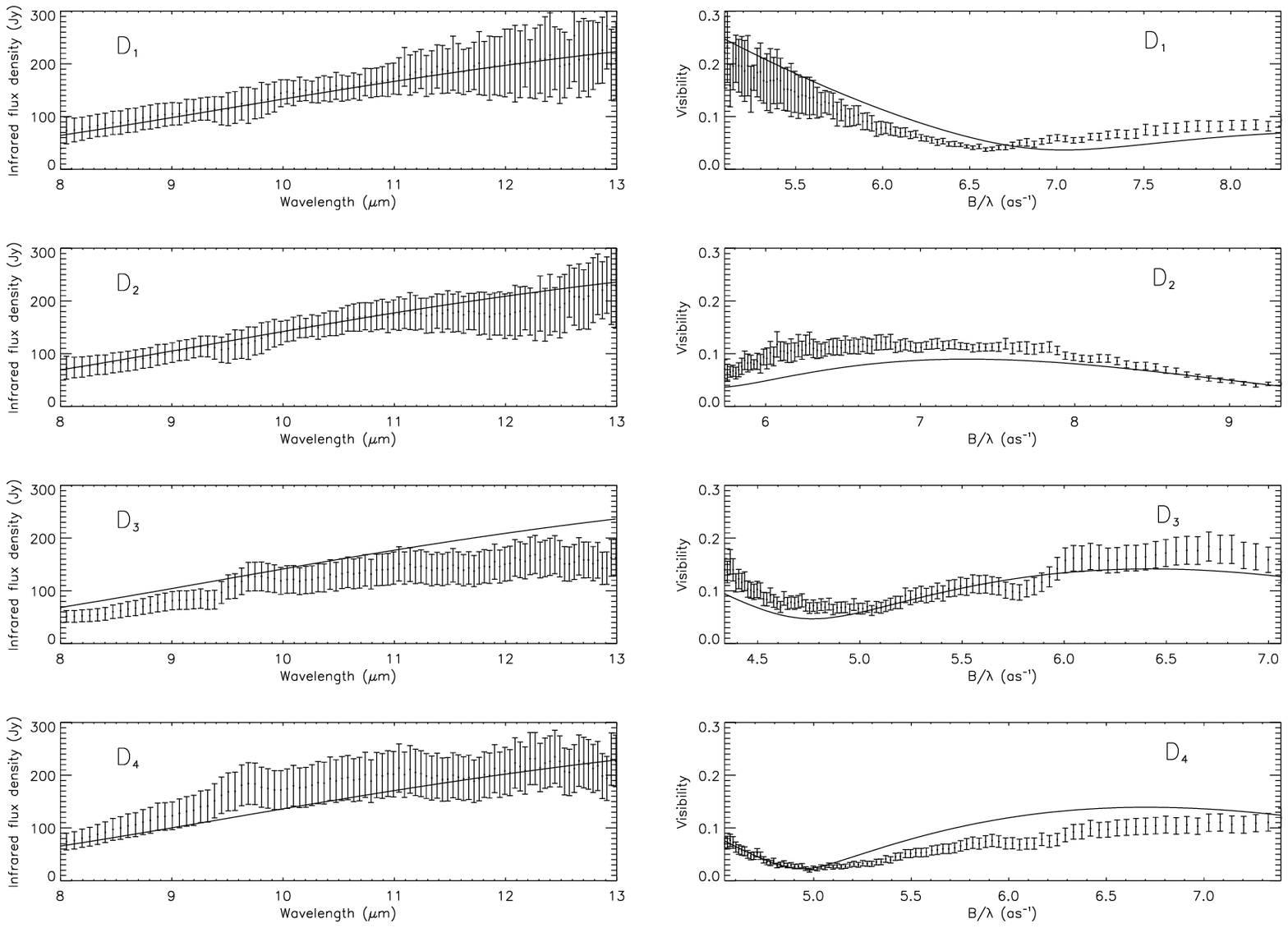}
 \end{center}
 \caption{}
 \label{fig:flux_vis_bestfit_concave}
\end{figure}

\newpage
\begin{figure}
\centering
\includegraphics[angle=90,width=110mm]{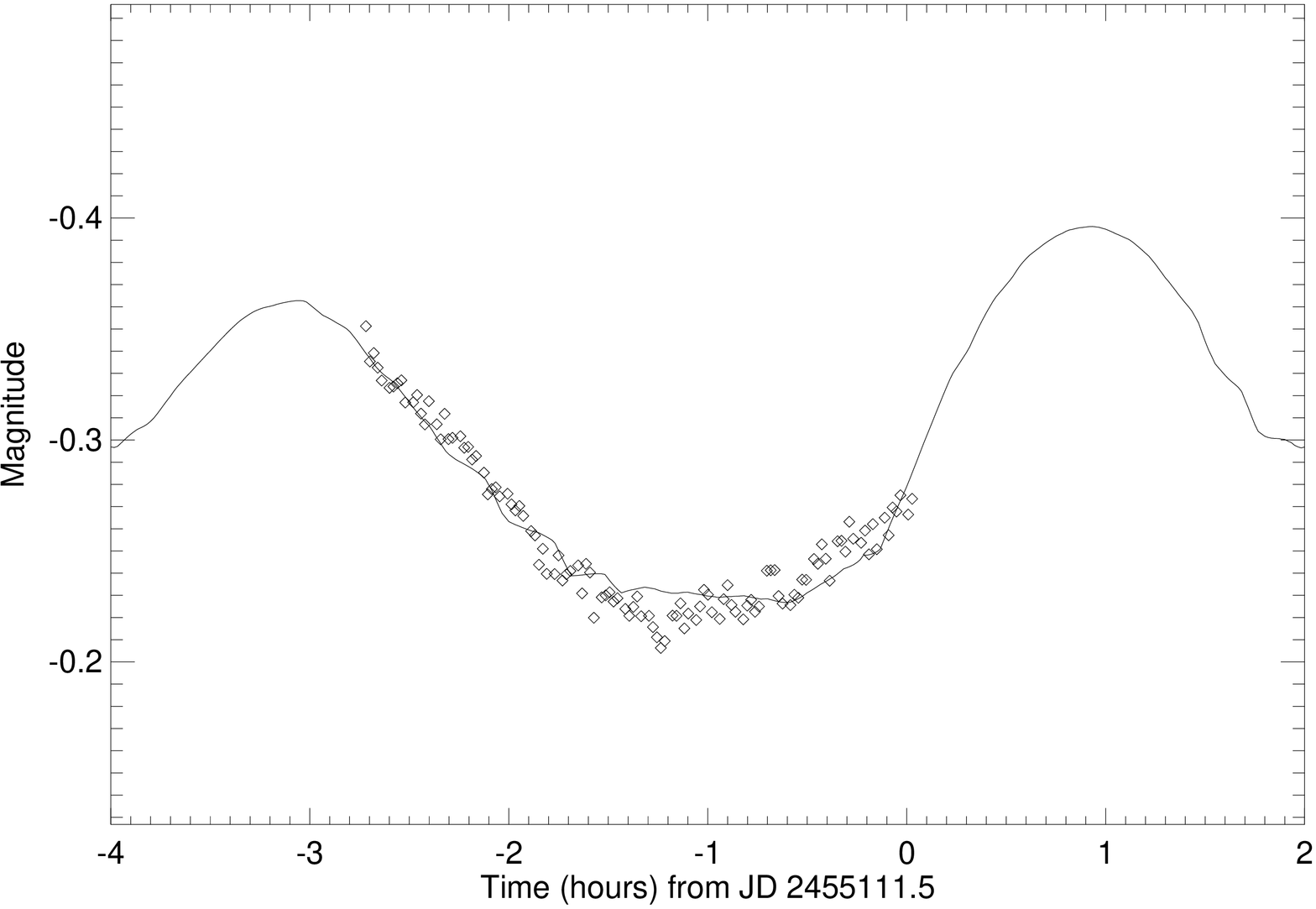}
\vspace{2mm}                                               
\caption{}
\label{fig:djamel}
\end{figure}

\newpage
\begin{figure}
\centering
 \includegraphics[width=130mm]{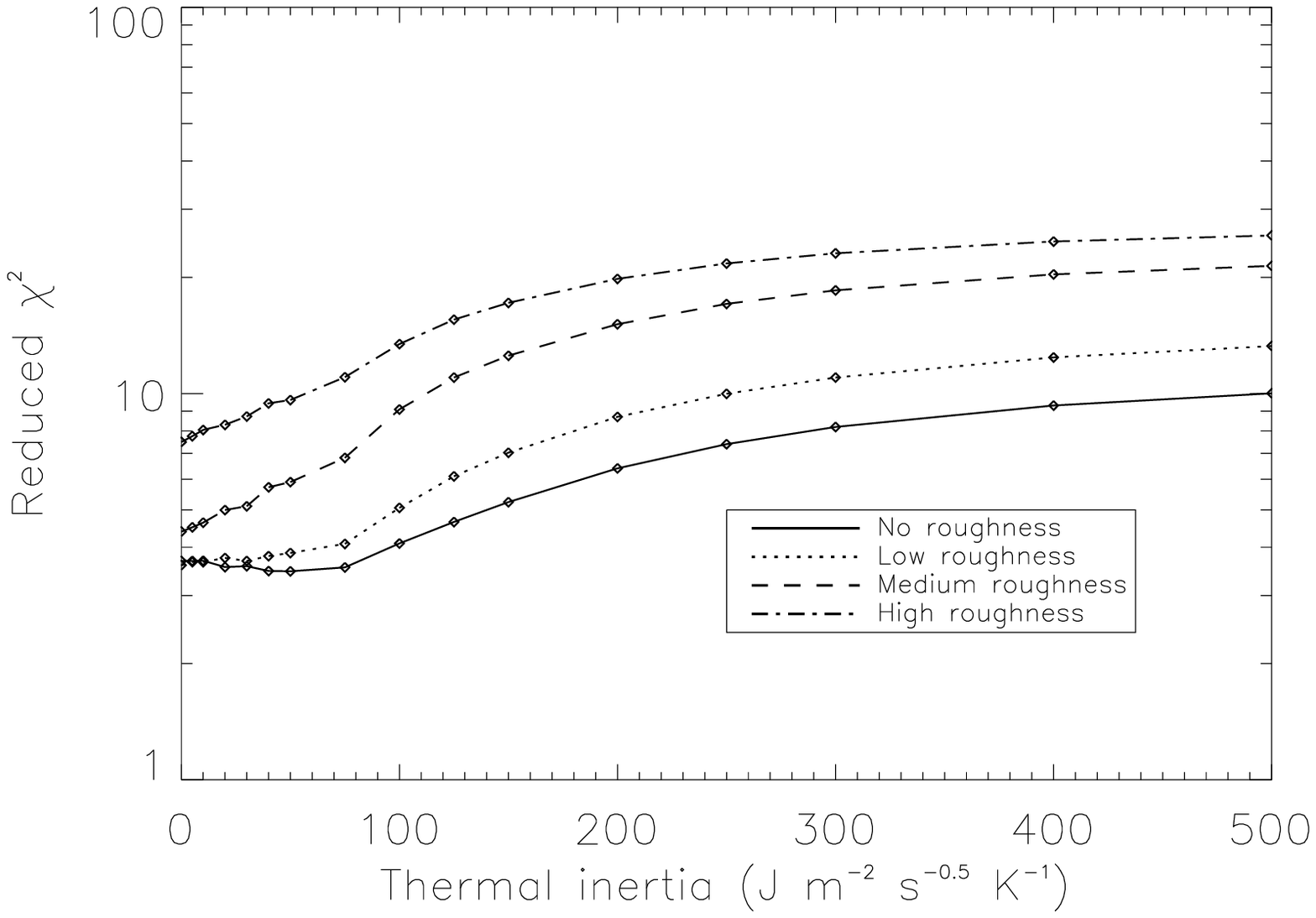}
 \caption{}
 \label{fig:chisq_convex}
\end{figure} 

\newpage
\begin{figure}
 \centering
 \includegraphics[width=130mm]{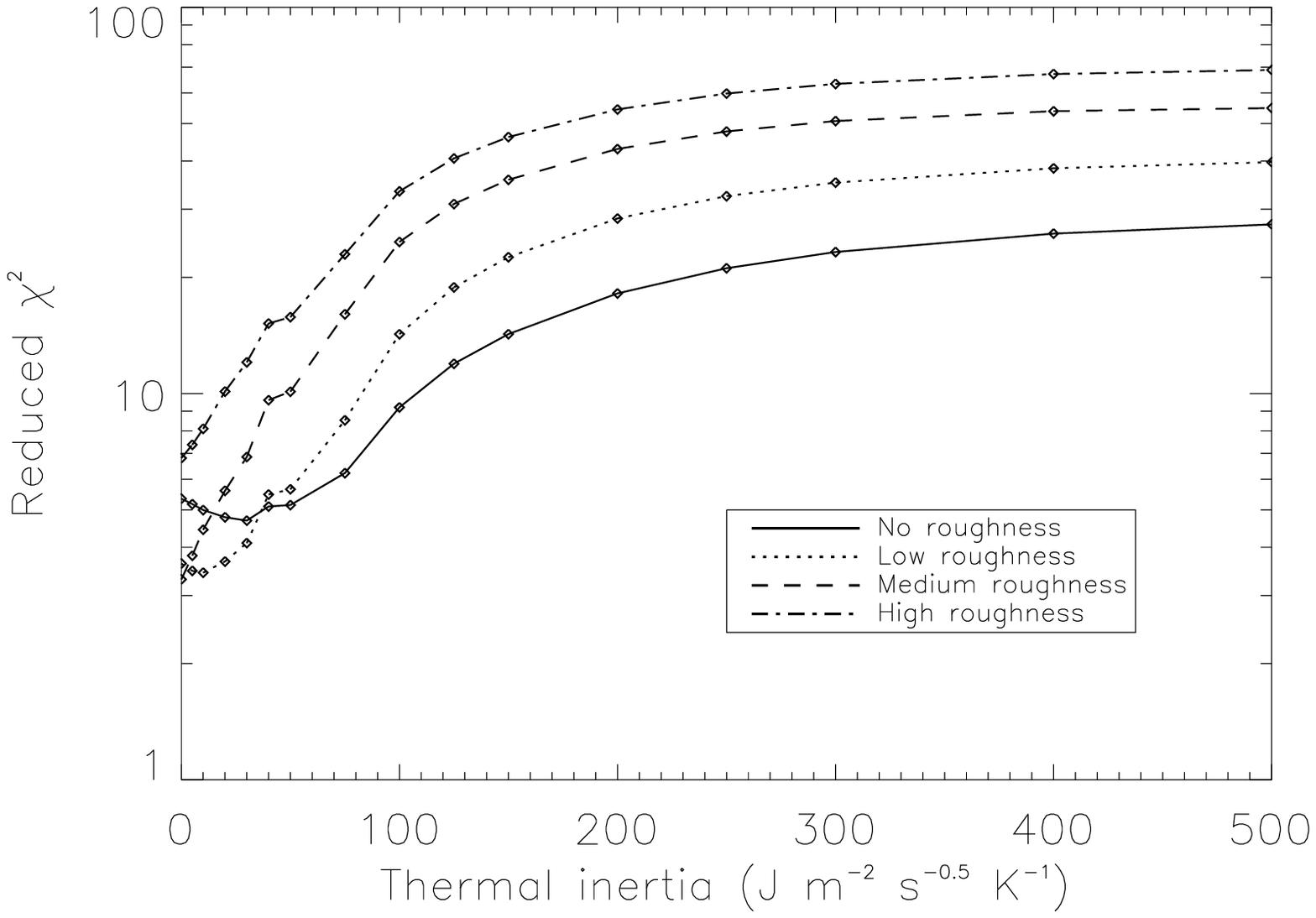}
 \caption{}
 \label{fig:chisq_concave}
\end{figure} 

\newpage
\begin{figure}
 \begin{center}
 \includegraphics[width=130mm]{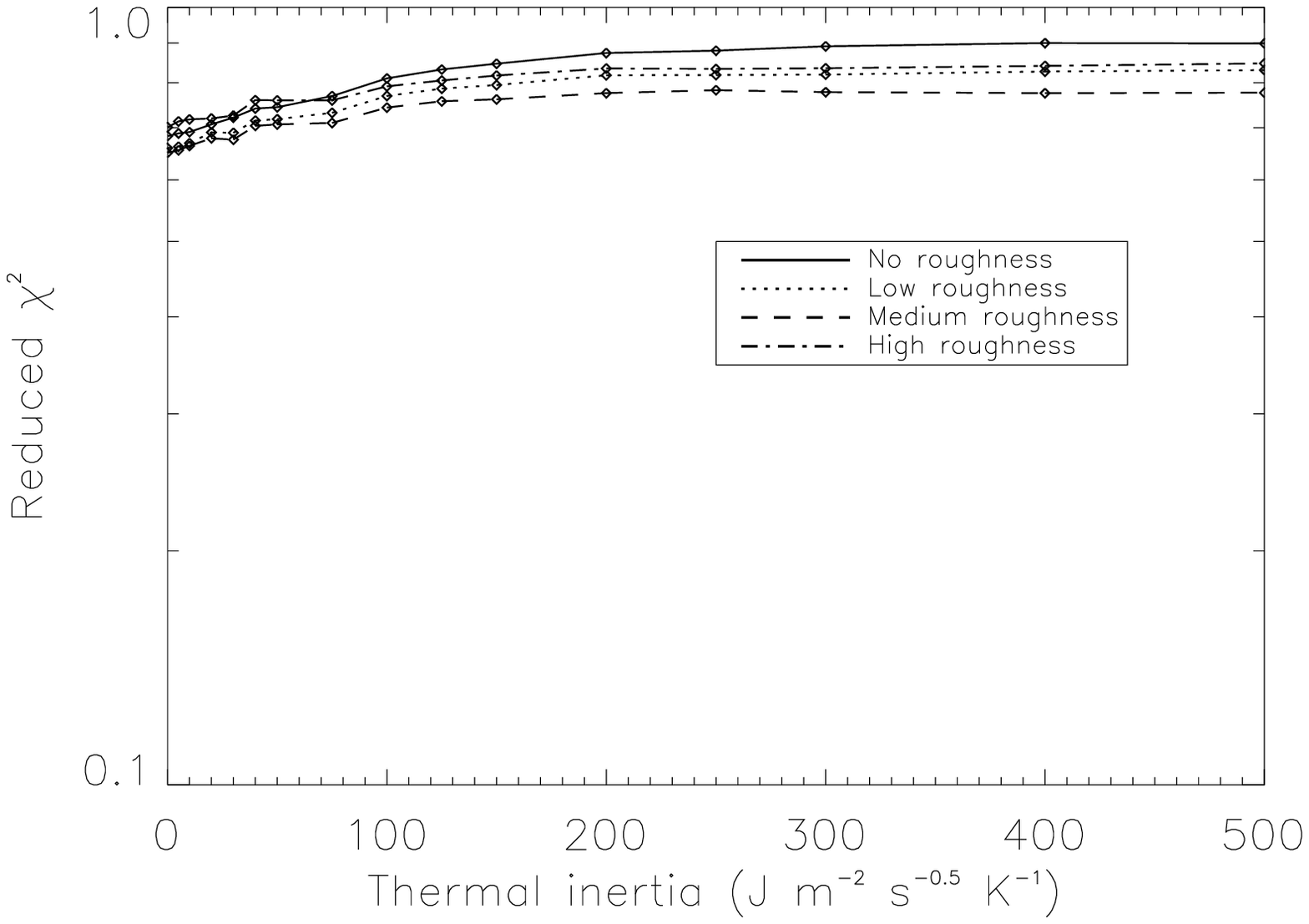}
 \end{center}
 \caption{}
 \label{fig:chisq_concave_flux}
\end{figure}

\end{document}